\documentclass[sigconf]{acmart}

\usepackage{graphicx}



\copyrightyear{2022}
\acmYear{2022}
\setcopyright{rightsretained}
\acmConference[CoNEXT-SW '22]{CoNEXT Student Workshop 2022}{December 9, 2022}{Roma, Italy}
\acmBooktitle{CoNEXT Student Workshop 2022 (CoNEXT-SW '22), December 9, 2022, Roma, Italy}
\acmDOI{10.1145/3565477.3569152}
\acmISBN{978-1-4503-9937-1/22/12}

\begin{document}

\title{Flow-based Network Intrusion Detection Based on BERT Masked Language Model}
\author{Loc Gia Nguyen}
\affiliation{
  \institution{Nagaoka University of Technology}
  \city{Nagaoka, Niigata}
  \country{Japan}
}
\email{s203145@stn.nagaokaut.ac.jp}
\author{Kohei Watabe}
\affiliation{
  \institution{Nagaoka University of Technology}
  \city{Nagaoka, Niigata}
  \country{Japan}
}
\email{k\_watabe@vos.nagaokaut.ac.jp}

\begin{abstract}
A Network Intrusion Detection System (NIDS) is an important tool that identifies
potential threats to a network. Recently, different flow-based NIDS designs
utilizing Machine Learning (ML) algorithms have been proposed as potential
solutions to detect intrusions efficiently. However, conventional ML-based
classifiers have not seen widespread adoption in the real-world due to their
poor domain adaptation capability. In this research, our goal is to explore the
possibility of improve the domain adaptation capability of NIDS. Our proposal
employs Natural Language Processing (NLP) techniques and Bidirectional Encoder
Representations from Transformers (BERT) framework. The proposed method achieved
positive results when tested on data from different domains.
\end{abstract}

\maketitle

\section{Introduction}
It is common in practical application of NIDS for there to be a change in the
data distribution between its training data and the data it encounters when
deployed. Conventional ML algorithms often adapt poorly to such change, which
limit their usefulness in real-world scenarios~\cite{survey}. To address this,
Energy-based Flow Classifier (EFC)~\cite{efc} was proposed as a solution.
Despite having good adaptability, EFC produces high false positives rate for
domains where the distribution of features of malicious flows overlap with that
of benign flows. We theorize that the reason for the limitations of conventional
ML algorithms and EFC is the use of singular flows as input data, as the
classifier can only model the distribution of features within a flow. This
limitation can be overcome with the use of sequences of flows, allowing the
classifier to further models the distribution of a flow in relation to other
flows. To utilize the context information from a sequence of flow, we use the
BERT framework, which is able to process inputs in relation to all the other
inputs in a sequence.

BERT\cite{bert} is a transformer-based machine learning technique for NLP
developed by Google. The BERT framework is comprised of two steps: pre-training
and fine-tuning. In pre-training, the BERT model is trained on unlabeled data.
For fine-tuning, the model is first initialized using the pre-trained
parameters, and then trained using labeled data from the downstream tasks. BERT
is pre-trained with two unsupervised tasks, which are Masked Language Modeling
(MLM) and Next Sentence Prediction (NSP). In MLM, some of the words in a
sentence are replaced with a different token. The objective is to predict the
original value of the masked words based on other unmasked words in the
sentence. In NSP, BERT takes sentence pairs as input. The objective is to
predict whether the second sentence in the pair is the next sentence in the
document. For fine-tuning, task-specific inputs and outputs are added to a
pre-trained BERT model.

Our research employs CIDDS-001\cite{cidds} and CIDDS-002\cite{cidds2} data sets
that contains flow samples from a small business environment emulated using
OpenStack. CIDDS-001 also contains real traffic flow samples captured from an
external server directly deployed on the internet.

\section{Proposal}
\begin{figure}
  \centering
  \includegraphics[width=\linewidth]{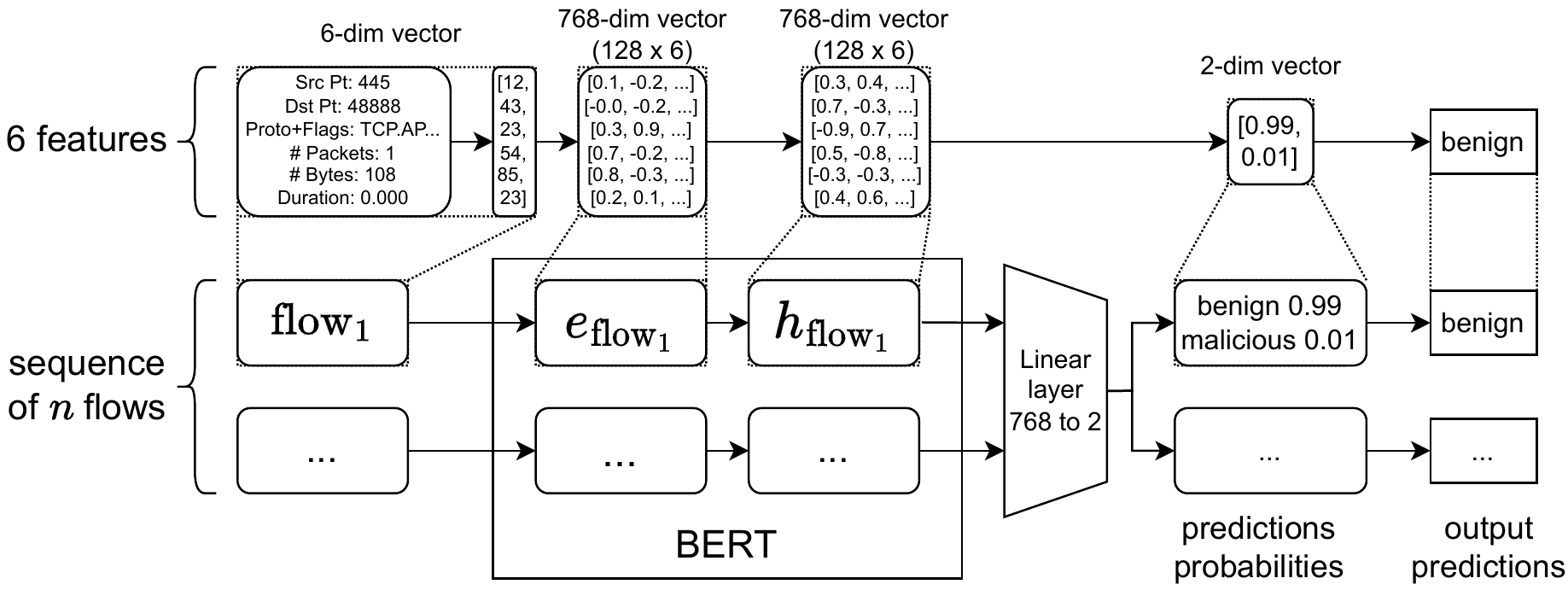}
  \caption{Proposed system architecture}
  \label{fig:prop}
\end{figure}

We first organize network traffic flows into  structures similar to a language,
treating a flow as a word and a sequence of flows as a sentence. In this study,
the BERT model is pre-trained with only the MLM task. For fine-tuning, a linear
layer with softmax output is used. It is important to preserve the distribution
of flows within a sequence; therefore, during training, the data set is not
shuffled. A training sample is generated by selecting a segment of flows from
the data set at random.

The overall architecture of the system is illustrated in Figure~\ref{fig:prop}.
Six features from each flow (Src Pt, Dst Pt, Proto+Flags, Packets, Bytes,
Duration) are used as input data, these features are discretized as described
in EFC's original paper~\cite{efc}. The discrete value of each features are
encoded as numbers~($\mathrm{flow}_i$). BERT decodes each number into a
128-dimension vector, concatenates them to form a 768-dimension
vector~($e_{\mathrm{flow}_i}$), and processes it to produce a different
768-dimension vector~($h_{\mathrm{flow}_i}$). The output of BERT is then passed
through a Multilayer Perceptron classifier (a linear layer with softmax output),
which reduces the dimension from 768 to 2. This 2-dimension vector represents
the predicted probability of the flow being benign and malicious (e.g. benign
0.99, malicious 0.01). The predicted class of the flow is the class with the
higher probability (e.g. benign).

We used data from three different domains: CIDDS-001 OpenStack, CIDDS-001
External Server, and CIDDS-002 to evaluate the domain adaptation capability of
the proposed method. Average composition of the data sets used in the experiment
are shown in Table~\ref{tab:data}. Training was performed on one set of
{\it CIDDS-001 large}. While testing was performed on {\it CIDDS-001 internal},
{\it CIDDS-001 external}, and {\it CIDDS-002}, each containing ten sets randomly
selected from the full data sets. This testing scheme mimics the one used by
Camila {\it et al.}~\cite{efc} to make the results of the proposed method more
comparable to those of EFC. Flows labeled $normal$ are considered benign, while
those labeled otherwise are considered malicious. For {\it CIDDS-001 external},
flows labeled $unknown$ and $suspicious$ are considered benign and malicious
respectively.

We assess the performance of our method in comparison to EFC and  ML classifiers
including Decision Tree (DT), K-Nearest Neighbors (KNN), Multilayer Perceptron
(MLP), Naive Bayes (NB), and Support Vector Machine (SVM). Each classifier's
performance is measured using Accuracy and F1 score~\cite{survey}.

\section{Results and Discussion}
\begin{table}
  \caption{Average composition of data sets}
  \label{tab:data}
  \resizebox{\linewidth}{!}{
    \begin{tabular}{lrclrclrclr}
    \toprule
    \multicolumn{5}{c}{CIDDS-001}
    &&\multicolumn{2}{c}{CIDDS-001}
    &&\multicolumn{2}{c}{CIDDS-002}\\
    \multicolumn{5}{c}{OpenStack}
    &&\multicolumn{2}{c}{External Server}
    &&\multicolumn{2}{c}{}\\
    \cline{1-5}\cline{7-8}\cline{10-11}
    \multicolumn{2}{c}{1 set}
    &&\multicolumn{2}{c}{10 sets}
    &&\multicolumn{2}{c}{10 sets}
    &&\multicolumn{2}{c}{10 sets}\\
    \multicolumn{2}{c}{CIDDS-001 large}
    &&\multicolumn{2}{c}{CIDDS-001 internal}
    &&\multicolumn{2}{c}{CIDDS-001 external}
    &&\multicolumn{2}{c}{CIDDS-002}\\
    \cline{1-2}\cline{4-5}\cline{7-8}\cline{10-11}
    label&\#&&label&\#&&label&\#&&label&\#\\
    \cline{1-2}\cline{4-5}\cline{7-8}\cline{10-11}
    $normal$&25178585&&$normal$&10000&&$unknown$&10000&&$normal$&10000\\
    $dos$&2775655&&$dos$&9000&&$suspicious$&10000&&$scan$&10000\\
    $portScan$&194642&&$portScan$&935&&&&\\
    $pingScan$&5266&&$pingScan$&45&&&&\\
    $bruteForce$&4992&&$bruteForce$&20&&&&\\
    \cline{1-2}\cline{4-5}\cline{7-8}\cline{10-11}
    Total&28159140&&Total&20000&&Total&20000&&Total&20000\\
    \bottomrule
    \end{tabular}
  }
\end{table}

Table \ref{tab:large} shows the average performance and standard error for each
classifier. All classifiers achieved higher Accuracy and F1 Score on
{\it CIDDS-001 internal} test sets (same domain as training data) compared to
the other test sets (different domains from training data). Both the proposed
method and EFC maintained performance across the two different domains, with
the proposed method outperforming EFC.

We also experimented with training the classifiers on smaller but balanced data
sets (containing 80000 flows with the same proportion of labels as in
{\it CIDDS-001 internal}). However, performance was worse for all classifiers
when compared to those trained on {\it CIDDS-001 large}. Notably, the
performance of the proposed method was significantly affected for all domains.
By creating balanced data sets, the distribution of flows within a sequence was
also altered. This suggests that the distribution of flows within a sequence is
learned by the model of the proposed method.

\begin{table}
  \caption{Average classification performance and standard error}
  \label{tab:large}
  \resizebox{\linewidth}{!}{
    \begin{tabular}{lcccccc}
    \toprule
    &\multicolumn{6}{c}{Train CIDDS-001 large}\\
    \cline{2-7}
    &\multicolumn{2}{c}{Test CIDDS-001 internal}
    &\multicolumn{2}{c}{Test CIDDS-001 external}
    &\multicolumn{2}{c}{Test CIDDS-002}\\
    \cline{2-7}
    Classifier&Accuracy&F1 score&Accuracy&F1 score&Accuracy&F1 score\\
    \hline
    Proposal&0.994(0.002)&0.994(0.002)&{\bf0.895(0.027)}
    &{\bf0.904(0.022)}&{\bf0.916(0.045)}&{\bf0.877(0.072)}\\
    EFC&0.941(0.005)&0.941(0.004)&0.747(0.039)
    &0.796(0.025)&0.846(0.046)&0.800(0.070)\\
    DT&{\bf0.996(0.001)}&{\bf0.996(0.001)}&0.870(0.028)
    &0.874(0.024)&0.818(0.061)&0.707(0.106)\\
    KNN&0.989(0.002)&0.989(0.003)&0.839(0.008)
    &0.811(0.011)&0.818(0.061)&0.707(0.106)\\
    MLP&0.992(0.001)&0.992(0.001)&0.573(0.009)
    &0.285(0.023)&0.832(0.059)&0.729(0.108)\\
    NB&0.903(0.002)&0.892(0.002)&0.500(0.000)
    &0.000(0.000)&0.499(0.000)&0.001(0.000)\\
    SVM&0.570(0.015)&0.245(0.047)&0.738(0.016)
    &0.718(0.021)&0.513(0.013)&0.090(0.037)\\
    \bottomrule
    \end{tabular}
  }
\end{table}

\section{Conclusion and Future Work}
In this study, we suggest the use of singular flows input to be a possible
explanation for the poor domain adaptation capability of conventional ML-based
classifiers. Then we proposed the used of sequences of flows to address this
limitation. We utilized BERT model for the representation of flow sequences
and an MLP classifier to discriminate between benign and malicious flows.
Early experimental results showed that the proposed method is capable of
achieving good and consistent results across different domains. However, more
extensive testing on recent data sets is needed to further evaluate its domain
adaptation capability.

In future work, we plan to investigate the impact of flow sequence sampling
method on result, such as using only benign flows or grouping flows into
sequences that originate from the same hosts. Making the system less reliant on
labeled data is another research goals of ours. We aim to achieve this by
modeling sequences of benign flows then look for anomalous representations
produce by BERT, indicating malicious flows.

\begin{acks}
This work was partly supported by JSPS KAKENHI Grant Number JP20H04172.
\end{acks}

\bibliographystyle{ACM-Reference-Format}
\bibliography{ref.bib}

\end{document}